\documentclass[9pt,onecolumn,twoside]{revtex4-2}

\usepackage{graphicx}
\usepackage{amsmath}
\begin{document}

\title{Active bacterial pattern formation in evaporating droplets}


\author{Twan J.S. Wilting$^1$}
\author{Adriana W.B.P. Reijnier$^1$}
\author{Michiel H.M. Brebels$^1$}
\author{Alexandre Villi{\'e}$^1$}
\author{R{\'e}my Colin$^2$}
\author{Hanneke Gelderblom$^{1,3}$}
\affiliation{$^1$ Department of Applied Physics and J. M. Burgers Center for Fluid Dynamics, Eindhoven University of Technology, P.O. Box 513, 5600 MB, Eindhoven, The Netherlands}
\affiliation{$^2$Max Planck Institute for Terrestrial Microbiology and Center for Synthetic Microbiology (SYNMIKRO), Karl-von-Frisch-strasse 10, 35043 Marburg, Germany}
\affiliation{$^3$Institute for Complex Molecular Systems, Eindhoven University of Technology, P.O. Box 513, 5600 MB, Eindhoven, The Netherlands}
\email[Corresponding author]{h.gelderblom@tue.nl}

\date{\today}
\begin{abstract}

Bacteria living on surfaces are often confined to droplets. When these droplets evaporate, the motion of the liquid-air interface and the associated internal capillary flow confine the bacteria. Here we study how \emph{E. coli} bacteria interact with this capillary confinement and agglomerate at the droplet's contact line. 
We identify three different types of bacterial pattern formation that depend on the bacterial activity and the environmental conditions imposed by the evaporating droplet. When the evaporation is fast, the bacteria are slow or the suspension is dilute, a uniform contact-line deposit forms. However, when the capillary confinement concentrates the bacteria at the contact line beyond a critical number density, localized collective motion spontaneously emerges. In that case, the bacteria induce a local stirring of the liquid that allows them to self-organize into periodic patterns and enables them to collectively escape from the contact line. At very high number densities, these periodic patterns get destabilized by bacterial turbulence in the bulk of the droplet resulting in the formation of mobile bacterial plumes at the contact line.
Our results show how the subtle interplay between the bacteria and the capillary flow inside the droplet that surrounds them governs their dispersal.
\end{abstract}

\maketitle

Bacteria living on surfaces are usually confined to liquid droplets \cite{Orevi:2021}. Common examples of these bacterial habitats are rain- or dew droplets on plant leaves \cite{Leveau:2019, Grinberg:2019, Gilet:2015}, waste-water droplets on contaminated surfaces \cite{Knowlton:2018,majee2021}, or respiratory droplets \cite{bourouiba2021}. Over time, these droplets evaporate, thereby confining and concentrating the bacteria. 
Motile bacteria at high number density show self-organized collective motion \cite{Dombrowski2004, Sokolov:2007, Colin:2019}. When dense bacterial suspensions interact with a confining boundary, a rich variety of phenomena emerges, such as spontaneous formation of directed flows \cite{Wioland:2016}, self-induced ordering \cite{lushi2014fluid}, vortex formation \cite{wioland2013confinement}, directed-swimming induced instabilities \cite{kasyap2012chemotaxis}, or the formation of interfacial protrusions \cite{Xu:2023}. In many of these studies, bacteria were confined by rigid walls, while confinement by the free surface of droplets \cite{Dombrowski2004, tuval2005, Xu:2023} has been much less studied. Moreover, in studies where bacteria are confined to droplets, evaporation is often deliberately suppressed to minimize the influence of evaporation-driven flows \cite{Dombrowski2004,tuval2005,Xu:2023}. In droplets containing passive colloidal particles, evaporation-driven capillary flows \cite{Gelderblom2022} are well known to induce the formation of ring-shaped stains \cite{Deegan:1997}. Only few studies \cite{nellimoottil2007, sempels2013, Kasyap2014, thokchom2014} focus on bacterial suspensions in evaporating droplets. In one of these works \cite{Kasyap2014}, the bacteria were found to self-organize at the contact line into radially inward pointing jets, which was hypothesized to be caused by the active stress of the swimming bacteria, but no mechanistic explanation for the phenomenon was demonstrated. Hence, how bacteria interact with the natural confinement and associated capillary flow imposed by an evaporating droplet has remained largely unexplored.

In this work, we show how the confinement by an evaporating droplet can cause \emph{E.~coli} bacteria to self-organize into different patterns, depending on the experimental conditions. 
To this end, we performed experiments with \emph{E.\,coli} confined to droplets evaporating from a glass substrate under controlled humidity conditions. With decreasing humidity, increasing swimming speed or bacterial number density, we identify three different bacterial patterns that may form: (I) the well-known coffee ring \cite{Deegan:1997}, (II) periodically spaced, radially-inward pointing fingers, as was observed before \cite{Kasyap2014}, (III) a novel pattern consisting of fluctuating fingers that move along the contact line, split, merge and undergo a sweeping motion. 
Systematically varying ambient humidity, bacterial number density and swimming speed, we mapped the full experimental phase diagram of these patterns. Combining experiments and theoretical modeling, we then identified the physical mechanisms driving the transitions between patterns. Through detailed imaging of the fluorescently labeled bacteria at the edge of the evaporating drop, combined with velocimetry measurements, we demonstrated for the first time that fingering is caused by collective swirls emerging from the active bacterial motility at high number density near the contact line. We developed a theoretical model to quantitatively explain the transition from uniform to fingering deposits, combining the increase in density at the contact line due to the evaporation-driven flow with this local transition to collective motion, which occurs above a critical number density \cite{Stenhammer:2017}. Finally, we show that the sweeping finger pattern is caused by collective motion developing in the bulk of the drop. The final stains that remain after complete evaporation of the droplet still bear traces of the structure of these collective-motion induced phenomena, demonstrating their long-term implications for the bacterial distribution on surfaces.

Our findings illustrate how capillary confinement strongly affects bacterial self-assembly and dispersal, and, vice versa, how bacteria may manipulate their accumulation and overcome interfacial flows through collective motion.

\section*{Results}
\subsection*{Bacterial fingering patterns emerge during drop evaporation}

\begin{figure}
\begin{turnpage}
 \includegraphics[width=\linewidth]{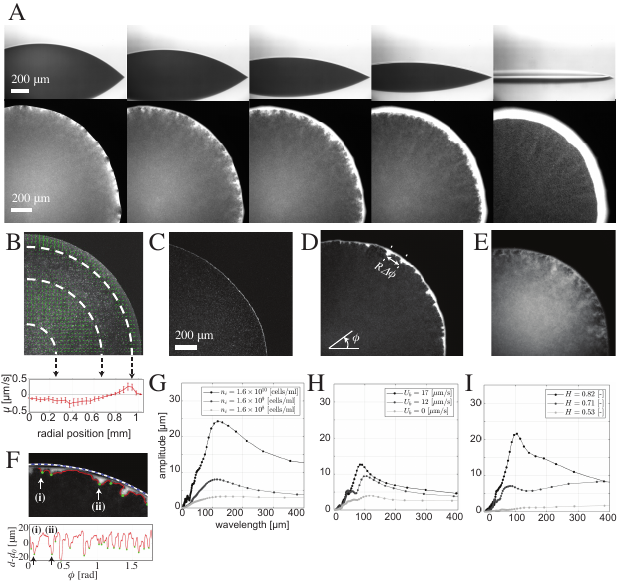}
    \caption{Pattern formation at the contact line of evaporating droplets containing motile \emph{E. coli} bacteria. (A) Top row: Side-view recordings the droplet's shape evolution. Bottom row: Bottom-view fluorescent microscopy images show the bacteria form radially inward pointing \emph{sweeping fingers} that move along the contact line, split and merge. (B) Fluid velocity field inside the droplet measured with 1.6-$\mu$m colloidal tracer particles by Particle Image Velocimetry (PIV) and time-averaged over 4 s. The corresponding radial velocity profile $u$ is shown in the bottom panel. Points are the mean and the error bar is the standard deviation over the azimuthal coordinate. (C-E) Snapshots of the three deposit types observed: (C) for an initial bacterial number density $n_i= 8.3\times 10^8$ cells/mL, swimming speed $U_b= 17~\mu$m/s and humidity $H=0.85$ a \emph{uniform deposit} forms. (D) For $n_i= 1.6\times 10^{9}$ cells/mL, $U_b=17~\mu$m/s and $H=0.53$ regularly spaced \emph{fingers} form at the contact line. The azimuthal coordinate $\phi$ is indicated. (E) For $n_i=1.6\times 10^{10} $ cells/mL, $U_b=17~\mu$m/s and $H=0.53$ \emph{sweeping fingers} form. (F) Top: zoom of the finger pattern with the interpolated and smoothed finger edge $d(\phi$)
    overlaid as a solid red line, and the droplet's contact line as a dashed blue-white line. Bottom: the edge of the finger structure $d-d_0$, where $d_0$ is the main radial position of the inner edge of the deposit, as a function of $\phi$. In both panels the same two fingers are marked by arrows and labeled (i) and (ii).
    (G-I) The Fourier spectra of the finger edge $d(\phi)-d_0$. Spectra for different values of (G) $n_i$ at fixed $U_b=17~\mu$m/s and $H=0.53$, (H) $U_b$ at fixed $n_i=6.4\times 10^9$ cells/mL and $H=0.53$, and (I) $H$ at fixed $n_i=1.9\times 10^9$ cells/mL and $U_b=17~\mu$m/s. The fingers have a robust wavelength of $\sim 100 \pm 24~\mu$m irrespective of the initial bacterial number density, swimming speed or humidity.}
    \label{fig:phenom}
    \end{turnpage}
\end{figure}
We let  1-$\mu$L sessile droplets of motility buffer containing motile \emph{E.\,coli} K12 bacteria deposited on Bovine-Serum-Albumin (BSA)-coated glass slides evaporate by maintaining the relative humidity ($H$) below $H=1$ inside a climate chamber (\emph{Materials and Methods}). The droplet evaporation and the pattern formation of the rod-shaped bacteria was recorded in side- and bottom-view (Fig.\,1 A). During evaporation, a radially outward capillary flow \cite{Deegan:1997} is present close to the contact line of the droplet. In the bulk of the droplet, an inward solutal Marangoni flow is present \cite{Marin2019}, driven by a surface tension gradient caused by the salts in the motility buffer; see Fig.\,\ref{fig:phenom} B. The strength of this Marangoni flow decreases in time \cite{Gelderblom2012}, and towards the end of the droplet life the entire flow field is directed radially outward (\emph{SI Appendix} Fig.\,S1). Over time, the outward capillary flow sweeps more and more bacteria towards the contact line. Initially, a deposition pattern with some irregularities forms at the contact line (first panel of Fig.~\ref{fig:phenom}A). These perturbations rapidly grow and evolve into radially inward pointing fingers (middle panels). These fingers are not stable: they move along the contact line and undergo a sweeping motion, while they split or merge (see movies S1 and S2). 

In control experiments with evaporating droplets of motility buffer that contain 1.6-$\mu$m colloidal particles or a mutant non-motile \emph{E.\,coli} strain (\emph{Materials and Methods}, \emph{SI Appendix} Fig.~S2 and movie S7) the sweeping finger pattern was not observed, and only a uniform  deposit was found, as in the classical coffee ring effect. Hence, the bacterial motility is key to the phenomenon observed. Moreover, in control experiments with motile bacteria but non-volatile droplets, the finger pattern was also absent and no contact-line deposit was formed. Hence, the evaporation-driven capillary flow is what causes the bacteria to reach the contact line, and the pattern formation is not driven by chemo- or aerotaxis \cite{kasyap2012chemotaxis, tuval2005} of the bacteria.

In the final instants of the droplet life an increasingly strong radially outward capillary flow exists \cite{marin2011order}. Moreover, in these final moments the bacteria stop swimming due to the increasing osmotic stress \cite{altendorf2009osmotic}; as the water evaporates from the motility buffer, the salt concentration increases and the bacteria eventually loose their motility. These two effects cause the sweeping motion to stop and the finger length to decrease. In these final instants, the bacteria get advected towards the contact line and rapidly fill the gaps between the fingers (see the last two panels of Fig.~\ref{fig:phenom}A). 
After complete dehydration of the droplet, the bacteria in the remaining stain are stuck and no longer move. However, the stain left after complete dehydration of the droplet still bears traces of the bacterial motility. Stains of non-motile bacteria (\emph{SI Appendix} Fig.\,S3 A) have a uniform width and the bacterial orientation is predominantly parallel to the contact line. By contrast, stains formed by motile bacteria show small ($\sim 5$ $\mu$m) undulations and contain domains of perpendicular bacterial orientation (\emph{SI Appendix} Fig.\,S3 B,C). These domains correspond to the former locations of the fingers. Thus, the phenomenology of finger formation leaves a permanent signature of bacterial motility in the final stain left by a bacterial droplet.

\subsection*{Phase diagram of the different patterns}
To identify under what conditions the finger pattern forms and construct the corresponding phase diagram, we performed a set of systematic experiments. In these experiments, we varied the initial number density $n_i$ of the bacteria, their swimming speed $U_b$ (by varying the age of the culture, see \emph{Materials and Methods}) and the evaporation rate of the droplet (by varying $H$). 

At the contact line, we observed three different types of bacterial patterns, depending on the experimental parameters: (I)  \emph{uniform} deposit (Fig.\,\ref{fig:phenom} C, movies S5 and S6), which only shows small irregularities of the order of a few bacterial lengths maximum. (II) A pattern of regularly-spaced, radially inward-pointing \emph{fingers} (Fig.\,\ref{fig:phenom} D and movies S3 and S4) with fixed azimuthal position (i.e.~small azimuthal variations remain within a finger width). (III) A pattern of long \emph{sweeping fingers} that show strong fluctuations in their azimuthal position over time and undergo a flapping motion (Fig.\,\ref{fig:phenom} E and movies S1 and S2). In contrast to the uniform deposit and the finger pattern, the sweeping fingers consist of bacteria that are moving around and have not yet deposited. 

We classified the type of deposit based on a quantitative analysis of the time evolution of the edge of the pattern $d(\phi,t)$. The deposit edge is indicated by the red line in Fig.\,1 F. The corresponding Fourier spectra (see \emph{Materials and Methods}) are shown in Fig.~\ref{fig:phenom} G-I. 
In our analysis, a pattern is classified as \emph{uniform}
when the amplitude of perturbations in the deposit edge remains smaller than $5~\mu$m, such as for $n_i=1.6\cdot10^8$ cells/mL in Fig.~\ref{fig:phenom} G. When the perturbations have an amplitude $> 5~\mu$m that persist for at least 10 s (to avoid influence of incidental short-lived spikes), and the perturbations are located at a fixed azimuthal position we classified them as \emph{fingers}.
For example, at $n_i=1.6\cdot10^9$ cells/mL in Fig.~\ref{fig:phenom} D,G we observe a finger-deposit with a perturbation amplitude of $\sim 8~\mu$m. 
When the azimuthal position of the fingers fluctuates along the deposit on length scales larger than the width of a finger it is called a \emph{sweeping finger} pattern. For example, Fig.~\ref{fig:phenom} E shows a sweeping finger pattern forms at $n_i=1.6\cdot10^{10}$ cells/mL. The corresponding Fourier spectrum (Fig.~\ref{fig:phenom} G) shows that the perturbation amplitude has increased to $\sim 24$ $\mu$m. At the large densities where sweeping fingers form, the consequent reduction in contrast of the pattern edge often made the determination of $d(\phi,t)$ very challenging. We therefore reclassified the patterns by hand to determine whether they show fingers or sweeping fingers. 

For each experimental condition ($n_i$, $H$, $U_b$), the amplitude spectrum quantifies the temporal evolution of the deposit profile -- initially uniform, from which fingers possibly emerge, which then possibly turn into sweeping fingers, before collapsing at the end of the evaporation process (Supplementary Fig.\,S4).
The most prominent finger pattern that emerges and its magnitude become more pronounced as $n_i$, $H$ or $U_b$ increases.
Indeed, the maximal pattern amplitude, quantified by the largest Fourier spectrum of the deposit edge that was observed over the course of an experiment, monotonously increases with $n_i$, $H$ and $U_b$ (Fig.\,\ref{fig:phenom} G-I). When fingering emerges, the Fourier spectrum always shows a broad peak, indicating a relatively wide range of possible finger wavelengths around a dominant one. This dominant wavelength of the (sweeping) finger pattern is always of the order of 100 $\pm 24\mu$m and robust to variations in $n_i$, $H$ and $U_b$.

\begin{figure}
\includegraphics[width=0.5\linewidth]{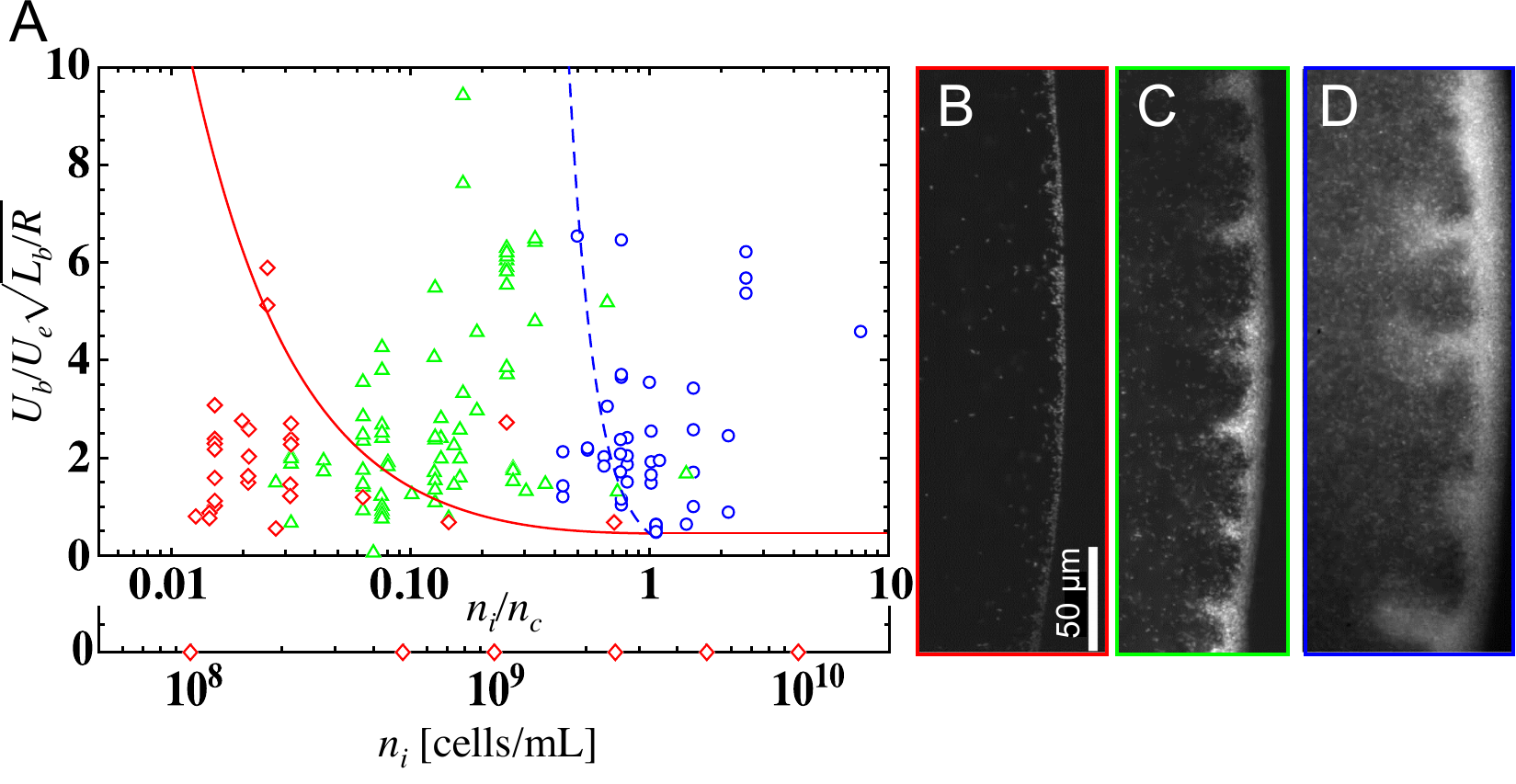}
\caption{(A) Phase diagram of the three different contact-line patterns formed by motile \emph{E. coli} bacteria: uniform (in red), fingers (in green), and sweeping fingers (in blue). On the $y$-axis, the bacterial swimming speed is rescaled by the evaporation-driven capillary flow speed $U_e$ and the ratio of bacterial length $L_b$ to droplet radius $R$, on the $x$-axis the bacterial number density is rescaled by the critical number density for collective motion $n_c$ (see main text). The results obtained for a non-motile strain, corresponding to $U_b=0$ (and hence $n_c$ is undefined), are denoted on a separate axis. The solid red line corresponds to the theoretical prediction Eq.~(\ref{eq:crit1}) for the transition from a uniform to a finger-pattern, the dashed blue line denotes the transition from fingers to sweeping fingers according to Eq.~(\ref{eq:crit2}), with free parameter $N=3$. (B-D) Examples of the corresponding patterns formed at $H\approx 0.4$: (B) a uniform pattern for $n_i\approx 5\times 10^7$ cells/mL. (C) A pattern with fingers at $n_i\approx 5\times 10^8$ cells/mL. (D) A pattern with sweeping fingers formed at $n_i\approx 5\times 10^9$ cells/mL. Corresponding movies can be found in the \emph{SI Appendix}. The common scale bar for images (B-D) denotes 50 $\mu$m.}
\label{fig:phase}
\end{figure}

We classified the most prominent pattern that emerges as function of $n_i$, $H$ and $U_b$ (\emph{SI Appendix} Fig.\,S5). These results can be mapped onto a two-dimensional phase diagram (Fig.\,\ref{fig:phase}A). The first non-dimensional control parameter is the swimming speed $U_b$, which is typically $\sim 17\mu$m/s, rescaled by the characteristic evaporation-driven capillary flow velocity close to the contact line, $U_e \sqrt{R/L_b}$, with \begin{equation}
    U_e=\frac{2D_{va}c_s(1-H)}{\rho_\ell R\theta_0},\label{eq:ue}
\end{equation}
where $D_{va}=2.46*10^{-5}$ m$^2$/s is the diffusion coefficient of water vapor in air, $c_s=2.32*10^{-2}$ kg/m$^3$ is the saturated vapor concentration, $\rho_\ell=997.5$ kg/m$^3$ the density of water and $\theta_0$ the initial contact angle of the droplet \cite{Gelderblom2022}. In our experiments, $U_e$ ranges between 0.2 and 2.5 $\mu$m/s. The factor $\sqrt{R/L_b}\approx 10$, with $L_b$ the bacterial length, comes from the fact that we consider the capillary flow speed in the region where the patterns form, which is within a few bacterial lengths distance from the contact line.

The second non-dimensional parameter is the bacterial number density $n_i$ rescaled by the critical number density for collective motion \cite{Stenhammer:2017}
\begin{equation}
n_c=\frac{5}{B\tau\kappa},\label{eq:nc}  
\end{equation}
where we take $B=1$ as a measure for the swimmer's nonsphericity, and $\kappa=c \ell U_b$ for the dipolar strength \cite{Skultety:2020}, where $\ell=1.9$ $\mu$m is the dipolar length and $c$ a length scale accounting for the viscous drag on a bacterium obtained from \cite{drescher2011fluid}. For the tumbling time we use the experimentally determined value $\tau=1.4$ s close to the solid surface, where the experiments are performed (see \emph{Materials and Methods} and \emph{SI Appendix} Fig.\,S8). We then obtain $n_c=6.6\times10^9$ cells/mL. 
Figure \ref{fig:phase} shows that when the bacterial swimming speed is zero or  weak compared to the capillary flow, the deposit remains uniform. For a finite $U_b/U_e\sqrt{L_b/R}$, the initial number density largely determines the type of deposit that forms. For low number densities the deposit is always uniform, regardless of $U_b/U_e \sqrt{L_b/R}$. 
Fingers are found at intermediate values of $n_i/n_c$ and sweeping fingers appear when the number density is increased to $n_i/n_c\gtrsim 1$. At these large number densities we observe collective motion of bacteria throughout the entire droplet, not only at the contact line. Note that the fuzzy transition zone between the fingers and sweeping fingers in the phase diagram is a consequence of the manual pattern classification in the sweeping-finger regime, as discussed above.

\subsection*{Mechanism of finger emergence}
\begin{figure}
\includegraphics[width=0.5 \linewidth]{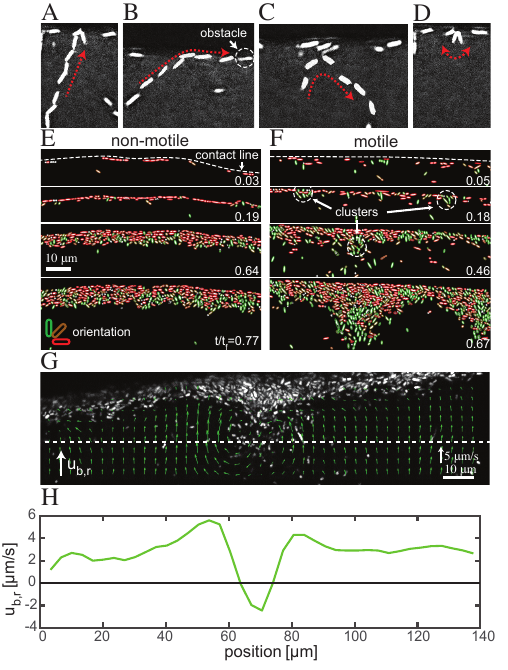}
\caption{Mechanism of finger nucleation and growth.
Top: 
Single-frame projections of image sequences showing the different fates of a bacterium swimming towards the contact line:
(A) The bacterium gets (temporarily) stuck at the contact line. (B) The bacterium re-orients and
continues to swim parallel to contact line until it reaches an obstacle, or (C) The bacterium re-orients and swims away from the contact line. Such reorientation either happens at the contact line itself or at cell bodies of other bacteria that are stuck there. (D) The orientation of a bacterium that is stuck at the contact line changes over time due to the bacterium's wiggling motion, which could cause it to escape from the deposit.
(E-F) Temporal evolution of bacterial deposits for (E) non-motile and (F) motile bacteria. Times $t/t_f$ are given in each panel, where $t_f$ is the droplet lifetime. The colors indicate the bacterial orientation, where green stands for perpendicular to the contact line and red for parallel. The bacterial orientation with respect to the contact line is obtained by fitting an ellipse to the cell body. Non-motile bacteria (E), once stuck, cannot escape the deposit and form a uniform pattern without any clusters or fingers. A deposit of motile bacteria (F) initially consists of clusters where bacteria nest into or escape from. Some of these clusters dissolve as bacteria escape, while others grow into fingers. (G) PIV measurements of motile bacteria around a finger. The velocity arrows represent a temporal average over 25 s. (H) Quantitative plot of the bacterial velocity in radial direction $u_{b,r}$ along a horizontal line (white dashed line in panel G) as function of the position, showing that bacteria escape the deposit at the fingers (resulting in $u_{b,r}<0$), while away from the fingers they move towards the deposit ($u_{b,r}>0$).}
\label{fig:onset}
\end{figure}

To gain insight into the physical mechanism that drives the fingering, we investigated the dynamics of finger growth in detail.
Using bottom-view recordings of the contact-line region at 40x and 100x magnification, we tracked the motion of individual bacteria and captured their interaction with the contact line. In Fig.\, \ref{fig:onset} (A-D) we observe that bacteria swimming towards the contact line can undergo three different fates: they either get (temporarily) stuck at the contact line (Fig.\,\ref{fig:onset} A), or re-orient themselves upon collision with the contact line and continue swimming parallel to it (Fig.\,\ref{fig:onset} B), or they make a sharp turn and swim away from the contact line towards the center of the droplet (Fig.\,\ref{fig:onset} C). These three types of behavior are well known \cite{berke2008,drescher2011fluid}, and observed in all experiments with motile bacteria, irrespective of their number density or the ambient humidity. 
The bacteria that get trapped at the contact line continue to rotate their flagella, which leads to a wiggling motion of their cell body \cite{hyon2012wiggling}, as shown in Fig.\,\ref{fig:onset} D. This wiggling sometimes causes a reorientation that allows the cell to swim away from the deposit \cite{Andac2019}. In addition, we observe that cells can reverse direction to escape from the deposit, presumably due to a reversal of the flagella rotation \cite{turner2010visualization} induced by the confinement in the liquid layer at the contact line.
Bacteria that remain stuck at the contact line become an obstacle for other bacteria approaching it. Figure \ref{fig:onset} B shows a track of a bacterium that swims parallel to the contact line towards such an obstacle. At the obstacle the bacterium gets trapped, causing the obstacle to grow. In the following, we define an obstacle of multiple bacteria at the contact line as a \emph{cluster}.
Note that not all bacteria that swim towards a cluster nest into it. Bacteria also get deflected by clusters, causing them to re-orient and swim away from the deposit; Fig.\,\ref{fig:onset} C.

In Fig.\,\ref{fig:onset} E and F we compare the growth of a deposit for a non-motile bacterial strain to that of a motile one. When non-motile bacteria arrive at the contact line, they are randomly oriented. The hydrodynamic forcing of the capillary flow aligns them parallel to the contact line \cite{dugyala2015evaporation}, where they remain stuck (Fig.\,\ref{fig:onset} E at time $t/t_f=0.03$, where $t_f$ is the lifetime of the droplet). Since these bacteria cannot swim, obstacles formed by one bacterium that is stuck at the contact line do not grow unless by chance another bacterium arrives at exactly the same spot. As a consequence, no clusters form, as shown in the next panels of Fig.\,\ref{fig:onset} E. At late times ($t/t_f=0.77$ and beyond), the capillary flow is so strong that the bacteria no longer have time to align and undergo an order-to-disorder transition, similar to what has been observed with colloidal particles \cite{marin2011order}.

Motile bacteria also arrive at the contact line with a wide range of possible orientations (Fig.\,\ref{fig:onset} F). 
The bacteria that arrive with an orientation perpendicular to the contact line get stuck, either temporarily until they manage to escape through wiggling their cell body, or permanently. Despite the hydrodynamic forcing by the background capillary flow, these trapped bacteria are able to maintain their perpendicular orientation
through a combination of hydrodynamic and steric interactions \cite{Baskaran:2009,Ezhilan:2013} and alignment with their neighbors. 
Indeed, in Fig.\,\ref{fig:onset} F the green colors at $t/t_f=0.05-0.18$ show that in a deposit formed by motile bacteria many of them have an orientation perpendicular to the contact line. 
Importantly, the polarity of these perpendicularly-oriented bacteria is such that their flagella are pointing away from the contact line, as inferred from their swimming direction before hitting the contact line (the flagella themselves are not visible in our recordings), otherwise they would be able to escape the deposit. A trapped pusher swimmer, such as \emph{E.~coli} will cause a Stokeslet disturbance of the fluid surrounding it that directs fluid away from the contact line \cite{drescher2009dancing}. Due to mass conservation, a secondary flow is established towards the sides of the swimmer, which results in a sideways attraction of nearby swimmers at the contact line, and thereby induces clustering \cite{thery2020self}. By contrast, bacteria with a parallel orientation do not get trapped but continue swimming along the contact line until they encounter an obstacle formed by perpendicularly oriented clusters of cells.

Clusters are mostly a transient phenomenon; as a result of cell reversals, and of their continuous re-positioning and re-orientation  \cite{Andac2019}, bacteria can escape the deposit and many of the initial clusters do not grow larger than one or two cell body lengths (see Fig.\,\ref{fig:onset} F at times $t/t_f=0.18-0.46$). 
In a cluster that does grow into a finger, we consistently observe that neighboring bacteria remain confined to circular trajectories (see movie S8). 
To further investigate this local accumulation of bacteria, we performed Particle Image Velocimetry (PIV) measurements of the bacterial velocity field (\emph{Materials and Methods}). Figure \ref{fig:onset} G shows the 25 s-averaged bacterial velocity field around a finger, where one clearly observes two circulation cells of approximately $\sim 20$ $\mu$m in diameter each. As quantified in Fig.\,\ref{fig:onset} H, at the finger, the average bacterial velocity field is directed radially inward, away from the contact line and \emph{against} the capillary flow. In between the fingers the flow is radially outward, towards the contact line. To further examine the motion of the background fluid, we added 360-nm passive colloidal tracer particles to the suspension (\emph{Materials and Methods}). Like the bacteria, these passive tracers were found to collect at the locations of the evolving fingers (\emph{SI Appendix} Fig.\,S6 A). 
Applying PIV to these tracers, we found that the background fluid undergoes a secondary circulation comparable to what was observed for the bacterial velocity field (\emph{SI Appendix} Fig.\,S6 B,C). 
Similar large-scale (i.e.\,beyond the scale of an individual swimmer) convection cells have previously been observed in experiments with aligned swimmers confined to a wall by an external magnetic field \cite{pierce2018hydrodynamic,thery2020self} or phototaxis \cite{drescher2009dancing}, and theoretically for biased swimming under the influence of a chemical attractant \cite{kasyap2012chemotaxis}. In our case, it is the evaporation-driven flow that locally accumulates the bacteria at the contact line and causes them to align with the flow direction \cite{Kasyap2014}. 

We deduce from these measurements that the clusters grow into fingers via a hydrodynamics-driven instability in number density \cite{Kasyap2014,kasyap2012chemotaxis}. The accumulation of bacteria at the contact line by the capillary flow allows the local number density to exceed the critical density for collective motion emergence, i.e.\,locally $n>n_c$. The bacteria in a cluster then induce a secondary flow that transports more bacteria into it, and allow it to grow into a finger.  If however the suspension is too dilute, cluster growth is too slow for fingers to form \cite{thery2020self} before the droplet has completely evaporated. 
Consistently with this reasoning, when a localized blob of large bacterial number density was observed to arrive at a previously uniform contact-line deposit (\emph{SI Appendix} Fig.\,S7), fingers only grew in the impact region where the local number density exceeded the threshold ($n>n_c$).

In addition to local accumulation such that $n>n_c$, the bacterial swimming speed also needs to be strong enough to overcome the evaporation-driven capillary flow speed and form the convection cells. As a consequence, no fingering is observed for the non-motile mutants. Moreover, towards the end of the droplet lifetime the evaporation-driven capillary flow diverges \cite{marin2011order}, so that it overcomes the collective bacterial swimming and flushes all bacteria towards the contact line, homogenizing the final deposit, as we observed in experiments for all initial bacterial number densities. However, if during this final stage, when the bacteria have already been aligned and accumulated at the contact line by the capillary flow, we suddenly halt the evaporation by covering the droplet with a cap, the fingers immediately re-emerge (movie S9). As soon as the cap is removed and the evaporation continues, the fingers disappear again (movie S10). 
These results reinforce our conclusion that the fingering results from a local accumulation and alignment of cells that allows cells to swim collectively against the evaporation-driven flow when both $n>n_c$ and $U_b>u$, where the capillary flow velocity close to the contact line scales as $u\sim U_e\sqrt{R/L_b}$.

\subsection*{Theoretical interpretation}
To interpret the different regimes in the experimental phase diagram (Fig.\,\ref{fig:phase}), we now formulate quantitative theoretical criteria for the two regime transitions: from a uniform to a finger pattern, and from fingers to sweeping fingers. We first define a criterion for the transition from a uniform deposit to a deposit with fingers, which corresponds to the two conditions discussed above being satisfied.
First, the local number density should be above the critical number density $n_c$ for collective motion in a large enough region around the cluster to prevent its dissolution by bacterial diffusion. Second, the bacterial swimming speed should be large enough to locally overcome the capillary flow and prevent the bacteria from being flushed towards the contact line. 

To quantify when these conditions are met, we model the evolution of the local bacterial number density inside the evaporating droplet using an advection-diffusion equation (\eqref{eq:n*}, see \emph{Materials and Methods}). For the advection speed we use the height-averaged radial flow velocity $u(r,t)$ inside the droplet (\eqref{eq:capflow}), in the case of vapor-diffusion limited evaporation \cite{Deegan:1997,marin2011order}, with $r$ the radial distance from the center of the droplet and $t$ the time. Velocity $u(r,t)$ increases both as a function of $t$ and of $r$, and it diverges at all times for $r\to R$, i.e.\,at the contact line, and over all space when $t\to t_f$, with 
\begin{equation}
t_f=\frac{\pi R} {8 U_e},\label{eq:tf}
\end{equation}
lifetime of the droplet and $U_e$ given by \eqref{eq:ue}. Hence, as the ambient humidity decreases, $U_e$ increases and the lifetime of the droplet goes down, and vice versa. Solving the equation assuming azimuthal symmetry yields the height-averaged bacterial number density $n(r,t)$ \cite{moore2021nascent}. Due to the evaporation-driven capillary flow inside the droplet, the number density $n(r,t)$ increases towards the contact line, and as a function of time, as expressed by Eq.\,(\ref{eq:n}). Clearly, as the bacterial number density increases beyond $n_c$, the suspension is no longer dilute and active bacterial stress causes large-scale circulations \cite{Dunkel:2013} that redistribute the bacteria. Such large-scale circulations are not captured by Eq.\,(\ref{eq:n}). However, we can use Eq.\,(\ref{eq:n}) to identify the regions in space and time where the local number density reaches $n_c$. 

From now on, we non-dimensionalize all number densities as $\hat{n}=n/n_i$, all lengths as $\hat{r}=r/R$, all velocities as $\hat{u}=u/U_e$, and all times as $\hat{t}=t/t_f$.
The non-dimensional cell number density $\hat{n}$ depends on $\hat{t}$, $\hat{r}$, and a single non-dimensional parameter, the solutal P\'eclet number:
\begin{equation}
  \mathrm{Pe} =\frac{RU_e}{D_b}=\frac{3R}{\tau U_b}\frac{U_e}{U_b},
  \label{eq:pec} 
\end{equation}
which compares the effects of advection by the evaporation-driven flow ($U_e$) and cell diffusion, expressed via the motile cell diffusion coefficient $D_b=\tfrac{\tau U_b^{2}}{3}$ \cite{Lauga2020}. In our experiments, Pe typically varies between 2.0 and 26.
Motile bacteria also undergo passive diffusion, which is however negligible, with $D\simeq 0.4~\mu$m$^2$/s for non-motile \emph{E. coli} \cite{Burriel:2024}, compared to active motion and reorientation with an effective diffusivity $D_b\sim 2.3\times 10^2$ $\mu$m$^2$/s.

According to our experimental observations, the first condition for clusters to evolve into fingers is $\hat{n}(\hat{r},\hat{t};{\rm Pe})>\hat{n}_c$ in a sufficiently large band next to the contact line at $\hat{r}=1$, say at least $N$ bacterial lengths wide, with $N\sim O(1)$. Hence, this contact-line region should at least run over $\hat{r}_c<\hat{r}<1$ with $\hat{r}_c=1-NL_b/R$. Since the bacterial number density increases towards the contact line, $\hat{n}(\hat{r}=\hat{r}_c)>\hat{n}_c$ is enough to fulfill this condition over the entire region. 

The second condition will be fulfilled only until the increasing capillary flow speed (\eqref{eq:capflow}) exceeds the bacterial swimming speed in the contact line region. We define the cross-over time $\hat{t}_c$ as the moment when $U_b/U_e=\hat{u}(\hat{r}_c,\hat{t}_c)$. Upon Taylor expanding Eq.\,(\ref{eq:capflow}) close to the contact line, we find 
\begin{equation}\hat{t}_c= \max\{1-\tfrac{\sqrt{2}}{\pi}\tfrac{U_e}{U_b}\tfrac{1}{\sqrt{1-\hat{r}_c}},0\}\,.\label{eq:tc}\end{equation} 
For all $\hat{t}>\hat{t}_c$ the capillary flow is too strong for bacteria to swim against it and the circulation cells vanish. Note that if $\hat{t}_c\to 0$, the bacterial motion is too weak to overcome the evaporation-driven flow during the entire droplet lifetime, either because the bacteria are too slow or because the evaporation is too fast, and the deposit remains uniform.

Since $\hat{n}$ increases as a function of $\hat{t}$, we hypothesize that the most developed pattern one can get for a given experiment happens at $\hat{t}_c$.
Using the two conditions for the bacterial number density and swimming speed, we can find the P\'eclet number for which $\hat{n}(\hat{r}_c,\hat{t}_c;\mathrm{Pe})=\hat{n}_c$ such that clusters do not diffuse away but can grow into fingers, at least at the threshold time $\hat{t}=\hat{t}_c$; beyond that critical moment, the capillary flow speed is too strong and the bacteria are washed towards the contact line. 

In practice, to find the transition between the uniform and finger regime in the phase diagram of Fig.\,\ref{fig:phase}, we use \eqref{eq:pec} and \eqref{eq:n} to calculate for each value of $U_b/U_e\sqrt{R/L_b}$ what the initial number density $n_i$ should be to satisfy 
\begin{eqnarray}\label{eq:crit1}
& &\hat{n}(\hat{r}_c,\hat{t}_c;\mathrm{Pe}) =\\ & &\hat{n}\left(1-\frac{NL_b}{R}, 1-\frac{\sqrt{2}}{\pi}\frac{U_e}{U_b}\sqrt{\frac{R}{NL_b}}; \frac{3R}{\tau U_b}\frac{U_e}{U_b}\right)=\hat{n}_c.\nonumber  
\end{eqnarray}
Hence, to find the transition we determine when $\hat{n}$ just reaches $\hat{n}_c$ at time $\hat{t}_c$, where this critical time depends on $U_b/U_e\sqrt{R/L_b}$ (see \eqref{eq:tc}).
In Fig.~\ref{fig:phase} we plot criterion \eqref{eq:crit1} with $N=3$ together with the experimental data, and find good agreement for the entire experimental range of $U_b/U_e \sqrt{L_b/R}$ and $n_i/n_c$. Note that the absence of finger formation for non-motile bacteria is also captured by the model, as in that case $n_c\to\infty$ and the condition $U_b/U_e>\hat{u}$ is never fulfilled (hence $\hat{t}_c=0$), resulting in a uniform deposition pattern. 

In the sweeping finger regime, the fingers at the contact line destabilize. In experiments, we observed this behavior whenever the initial number density was so large that collective motion and large-scale circulations were not only found at the contact line, but also in the bulk of the droplet. We thus expect that these bulk turbulence patterns lead to the destabilization of the fingers and cause them to undergo a sweeping motion.
To determine the transition from fingers to sweeping fingers, we therefore calculate for what initial number densities large-scale collective motion occurs in the entire droplet (including the center $\hat{r}=0$) at $\hat{t}=\hat{t}_c$, the moment until when fingers can form. Hence, we determine for what $n_i$ we get
\begin{eqnarray}\label{eq:crit2}
& &\hat{n}(\hat{r}=0,\hat{t}_c;\mathrm{Pe}) =\\ & &\hat{n}\left(0, 1-\frac{\sqrt{2}}{\pi}\frac{U_e}{U_b}\sqrt{\frac{R}{NL_b}}; \frac{3R}{\tau U_b}\frac{U_e}{U_b}\right)=\hat{n}_c.\nonumber  
\end{eqnarray}
In Fig.~\ref{fig:phase} this transition criterion is denoted by the blue dashed line. Indeed, for $n_i/n_c>1/\hat{n}(0,\hat{t}_c;\rm{Pe})$ we only find the sweeping finger pattern in our experiments, whereas when $n_i/n_c$ decreases below this threshold the system transitions to the stable finger pattern. 

\section*{Discussion}
In summary, we found that motile bacteria that are confined to an evaporating droplet can form three different contact-line patterns, and we identified the control parameters and the underlying physical mechanisms governing the transition between pattern types. A uniform deposit forms when the swimming velocity of individual bacteria is too slow, the suspension is too dilute to allow for collective motion, or the evaporation and corresponding capillary flow is so fast that the bacteria are just swept to the contact line, as in the classical coffee-ring effect.
Periodic finger patterns in which bacteria collectively escape the contact line form when capillary confinement induces localized collective motion; the evaporation-driven flow aligns and accumulates bacteria at the contact line in such high densities that circulation cells appear, trapping bacteria around growing clusters. In experiments with an initial bacterial number density above the critical density for collective motion, a bulk instability causes convection cells in the entire droplet. These convection cells destabilize the fingers growing at the contact line, and a sweeping-finger deposit is formed.

Even though the finger formation appears to be a transient phenomenon, the final stain shows signatures of bacterial motility. Domains of perpendicular bacterial orientation persist in the final drying stain of a droplet that contained motile bacteria, and are absent in stains left by droplets containing non-motile cells. This anisotropy in bacterial orientation inside a deposit is important, as it is known to influence swarming and survival of bacteria on surfaces \cite{richard2020,majee2021}. Moreover, such anisotropy could alter drying crack formation, which plays a key role in many biological, environmental and industrial processes \cite{ma2022crack}, and could potentially be used for diagnostic purposes \cite{Sefiane:2010, Trantum:2012}.

The fingering pattern formed during droplet evaporation provides the bacteria with a means to collectively alter their dispersal on a surface. The contact line presents a harsh environment to the bacteria \cite{majee2021} due to the rapid dehydration, high osmotic pressure and large shear stresses. Therefore, a collective escape might be a viable method for bacteria to survive under such harsh conditions \cite{Orevi:2021}.
A number of pioneering studies have shown that bacteria are able to actively manipulate their dispersal in interfacial flows by e.g. biosurfactant secretion to change the bursting dynamics of a
bubble \cite{Poulain:2018}, to induce Marangoni flows \cite{Angelini:2009, sempels2013} or
to depin and move a sessile droplet \cite{hennes2017active}.
The present study shows that collective motion is yet another means for bacteria to actively influence their dispersal onto surfaces, by interacting with the droplet that confines them. How such interplay between motile bacteria and interfacial flow affects their future survival remains to be explored.


\section*{Materials and methods}
\subsection*{Preparation of bacterial suspension droplets}
The bacteria used in this work are rod-shaped, motile \emph{E.\,coli} K12 (strain MG1665, $\Delta$\emph{flu}, in which the major surface adhesin of \emph{E.\,coli} is deleted to prevent cell aggregation) and a non-motile ($\Delta$\emph{motAB}) mutant. The bacteria are fluorescently labeled via a Trc99a vector plasmid carrying an isopropyl $\beta$-D-thio-galactoside (IPTG) inducible promoter that expresses green fluorescent protein (GFP). 

The bacteria cultures are grown overnight in tryptone broth (TB; 1\% tryptone, 0.5\% NaCl,
pH 7.0) at $37^\circ$C  and 250 revolutions per minute in a rotary incubator followed by a 100x dilution into fresh TB the next day. The diluted culture is then allowed to grow in the rotary incubator until the bacteria reach the mid-exponential growth phase (suspension optical density of 0.6), to maximize swimming speed. The bacteria are harvested by centrifuging at 3600rpm for 5 minutes followed by two washing cycles where the bacteria are re-suspended into a motility buffer (MB; 3.915 g/L NaCl, 1.07 g/L K\textsubscript{2}HPO\textsubscript{4}, 0.52 g/L KH\textsubscript{2}PO\textsubscript{4} and 0.037 g/L EDTA), in which  they no longer proliferate and their number density remains constant. 

The bacterial number density obtained through this procedure is typically $n_i\approx(10\pm0.3)\cdot10^{10}$ cells/mL, which is then further diluted to the initial number density required for each experiment. From fluorescence-microscopy images at 100x magnification, we measured the bacterial body length to be $L_b\approx 1.75\pm 0.67$ $\mu$m and diameter $d_b\approx0.75\pm 0.15$ $\mu$m. The flagellar bundle is not fluorescent and therefore not visible but has an estimated typical length of $8$ $\mu$m \cite{turner2010visualization,turner2012growth}. The average swimming speed of a fresh culture is measured to be $U_b=16\pm 1$ $\mu$m/s in the bulk and $18\pm 1$ $\mu$m/s at the solid surface; see \emph{SI Appendix} Fig.\,S8.
As the culture ages, the swimming speed decreases \cite{SchwarzLinek2016}, which allows us to vary the bacterial swimming speed by using cultures of different age. For our cultures, the swimming speed decreased from $\sim 17$ $\mu$m/s to $\sim 10$ $\mu$m/s in 17 hours. In the experimental results, we used this linear relation between age of the culture and swimming speed to estimate the actual swimming speed of the bacteria and used these values to determine the phase diagram shown in Fig.\,\ref{fig:phase}. The bacteria swim by alternating runs and tumbles, with a mean run duration, determined by particle tracking, of 0.6~s in the bulk and 1.0~s at the surface, and a tumbling rate of  1.0-0.72 tumbles.s$^{-1}$ (bulk-surface); see \emph{SI Appendix} Fig.\,S8. Particle tracking was performed in MB supplemented with 0.01\% tween80 as surfactant preventing cell adhesion to glass, in a non-evaporating drop sandwiched between two glass slides that are about 150\,$\mu$m apart, close to the top slide (surface), and at mid-height between the slides (bulk), using the ParticleTracking2 ImageJ plugin (https://gitlab.gwdg.de/remy.colin/particletracking2).

To prevent adhesion of bacteria to the glass microscope slides used for our experiments, the glass slides are coated with 0.2 m\% Bovine Serum Albumin (BSA). We found that the BSA film was not always homogeneously deposited, which led to differences in substrate wettability and droplet shape \cite{bialopiotrowicz2001}. We therefore only used the substrates on which the droplet's contact line was circular. Evaporating the droplets from bare glass without BSA coating did result in finger patterns, but the tracking of bacteria was problematic due to their adherence to the glass.

\subsection*{Experimental set-up and imaging}
Sessile \emph{E.~coli} suspension droplets of $1 \mu$L are evaporated inside a custom-made transparent climate chamber that allows us to measure and control the relative ambient humidity and minimize disturbances in the surrounding airflow in the lab. Humidity control occurs via a humidity-temperature sensor (H-sensor) that is coupled to a controlled inflow of dry and wet air \cite{Boulogne:2019}. In this way, variations in relative humidity during an experiment are limited to a maximum of $5\%$. The temperature is not controlled but monitored throughout all experiments and shows variations of less than $0.5^\circ$C.

The droplet shape during evaporation is recorded with a side-view camera (Basler, acA2040-90um) operated at 1~frame/s and a 8x telecentric lens. The resulting field of view (FOV) is 1.4 x 1.4 mm$^2$, in which half the droplet is captured. We determine the contact angle $\theta(t)$ and droplet base radius $R(t)$ over time by fitting a spherical cap to the droplet contour. The side-view recordings are also used to determine the relative humidity $H$ in the far field of the droplet, through a fit of the relation \cite{Gelderblom2022}
\begin{equation}
    \theta(t)=\theta_0\left(1-t/t_f\right),\label{eq:theta}
\end{equation}
where $t_f$ depends on $H$ via \eqref{eq:tf} and \eqref{eq:ue}.
 From this analysis, we systematically find that relative humidity measured by the H-sensor is lower than the value of $H$ determined from the fit. As similar observation was reported for a different set-up \cite{seyfert2021evaporation}. We expect the difference between the measured and fitted $H$ to be caused by e.g.\,evaporative cooling that could locally decrease $c_s$ and thereby lower the evaporation rate \cite{Schofield:2018}.
In our results, we therefore always use the fitted value of $H$, and hence the measured rate of droplet evaporation to estimate the fluid flows within the droplet.

To measure the bacterial pattern formation, a bottom-view sCMOS camera (PCO, Panda 4.2, pixel size $6.5\times 6.5$ $\mu$m$^2$) with different objectives (Olympus PLN 10x, 40x and 100x) is used. At 10x magnification, the FOV is $1.14 \times 1.14$ mm$^2$, which allows us to measure one quarter of the droplet footprint. At this magnification, the inner and outer edge of the bacterial deposit are determined from the intensity of the fluorescent signal. The time-evolution of the deposit edge recorded at 1 frame/s (see figure \ref{fig:phenom} F) is used to classify the different types of deposit (uniform, finger, sweeping finger). For more details on the contact-line and visualization of individual bacteria, we used the 40x and 100x objectives, with corresponding FOVs of $0.28\times 0.28$ mm$^2$ and $0.116\times 0.116$ mm$^2$, respectively.

\subsection*{Particle Velocimetry measurements}
For the particle image velocimetry (PIV) and particle tracking velocimetry (PTV) experiments of the background fluid flow (in absence of the bacteria) we used naturally buoyant red-fluorescent polystyrene colloidal tracer particles with a diameter of $d_c=1.61\pm0.04$ $\mu$m and density $\rho_c=1.05$ kg/m$^{3}$ (PS-FluoRed, obtained via microParticles.de).  Recordings at 5-10 frames/s are made, at 10x or 40x magnification. For PIV, the data is pre-processed using custom-made Matlab code and the software PIVview3C (PIV\emph{TEC} GmbH). PIV interrogation windows of $71\times 71 $ $\mu$m$^2$ with 50\% overlap are used. The resulting velocity fields are time-averaged over 4 s. For PTV analysis, the DecofusTracker software \cite{barnkob2021} is used. To reduce the influence of Brownian motion and inaccuracies in particle detection, tracks are smoothed by a moving Gaussian average filter over 10 neighboring points.  

For the PIV experiments with the bacteria, we used a confocal microscope (Nikon, CFI Plan Apo VC) with 58-frames/s recordings and a 100x magnification objective, resulting in an FOV of $143\times 36$ $\mu$m$^2$. To obtain the average bacterial velocity field near a finger, as shown in Fig.~\ref{fig:onset} G, we time-averaged the displacement data over 10-100s. In the PIV analysis, interrogation windows of $7\times 7 $ $\mu$m$^2$ with 50\% overlap are used to obtain a $41\times 9$ grid of vectors. 

For the PIV experiments with mixtures of bacteria and colloids we used custom-made neutrally buoyant fluorescent colloidal particles (core: 2,2,2-trifluoroethyl methacrylate (TFEMA) with 1\% crosslinker ethylene glycol dimethacrylate (EGDMA) containing Rhodamine B-methacrylate as fluorescent dye).  The particles are coated with a non-adhesive shell consisting of polyethylene glycol monomethyl ether methacrylate (PEGMA, mw $\sim$ 500 g/mol) with 9\% crosslinker EGDMA \cite{kodger2015precise} to minimize the interaction between the bacteria and the colloids. The total particle diameter (core plus coating) is 360 nm.
A colloidal seeding density of 0.1\%w/w is used, which is large enough to achieve a sufficient tracer density close to the fingers (in particular since only time-averaged velocity fields were determined), but small enough to prevent interference with the finger formation by the bacteria. 

\subsection*{Wavelength analysis of the finger pattern}
To quantitatively analyze each type of deposit we tracked the deposit edge at the moment when the perturbations are largest and analyzed its Fourier spectrum (Fig.\,\ref{fig:phenom} F-I). To this end, we first binarized the images using adaptive thresholding and removed rough edges using a 20×20 pixel kernel. To find the approximate droplet center, a circle is fitted through the white pixels forming the deposit. To determine the inner (outer) edge of the deposit we then found the position where the intensity jumps from black to white (white to black) while moving radially outward from the approximate droplet center. The outer deposit edge is then used to again fit a circle, as is indicated by the dashed white-blue line in Fig.~\ref{fig:phenom} F, from which we find the exact position of the droplet center. The position of the inner deposit edge is then expressed in polar coordinates $\left(\phi, d(\phi)\right)$, with $d$ the radial coordinate of the edge and $\phi$ the azimuthal coordinate with respect to the droplet's center, at a resolution of $\tfrac{\pi}{2000}$. 
The resulting signal is shown in the bottom panel of Fig.~\ref{fig:phenom} F, where we have subtracted the mean inner edge location $d_0$ to center the signal around zero. We then calculated the Fourier spectrum of $d(\phi)-d_0$ using the Matlab Fast Fourier Transform algorithm to find the azimuthal period $\Delta\phi$ of the perturbations in the deposit edge and their amplitude. The corresponding wavelengths are then given by $\lambda=R\Delta\phi$, with $R$ the droplet radius. The spectra obtained are smoothed by a Gaussian average filter (window size of 10 data points) to remove noise and obtain the mean distribution.

\subsection*{Modeling of the local bacterial number density distribution}
To model their local number density distribution, we assume that the bacteria are initially homogeneously distributed throughout the droplet and have a random orientation and swimming direction.  Furthermore, we assume that the bacteria move in an unbiased way with a run-and-tumble motion that gives rise to a diffusive flux with an effective diffusivity $D_b$. We hence neglect any aerotactic bias \cite{tuval2005}, which is justified because in experiments with non-volatile droplets ($H=1$), the bacterial number density distributions remained homogeneous. 

On top of the diffusive transport, the evaporation-driven capillary flow causes a net drift of bacteria towards the contact line. 
Since our bacterial suspensions are initially dilute $(n_i\ll n_c)$, this capillary flow is not influenced by the presence of the bacteria, and we initially neglect active stress contributions to the flow. Clearly, these assumptions no longer hold at high bacterial densities; i.e.~directly at the contact line or in experiments where the initial number density is already beyond the critical number density for collective motion. 

In the dilute limit, the droplet-height averaged bacterial number density distribution $n(r,t)$,  where $r$ is the radial distance from the droplet center and $t$ is the time, is governed by the advection-diffusion equation \cite{moore2021nascent}
\begin{equation}
    \frac{\partial }{\partial t}\left(hn\right) + \frac{1}{r}\frac{\partial}{\partial r} \left[r  hn u - D_b r h \frac{\partial n}{\partial r}  \right] = 0,
    \label{eq:n*}
\end{equation}
where $h$ is the droplet height and $u$ the height-averaged evaporation-driven capillary flow velocity.
To calculate $u$ we follow the steps detailed in \cite{Gelderblom2022}, and assume that the evaporation rate of the droplet is limited by the diffusive transport of vapor in the surrounding (ambient) air \cite{Deegan:1997}, and that natural convection caused by the density difference between dry and humid air is negligible for the droplet sizes considered \cite{shahidzadeh-bonn_rafaï_azouni_bonn_2006}. Furthermore, the evaporation is assumed to occur in a quasi-steady fashion, as the timescale for vapor diffusion $t_D=R^2/D_{va} = O(10^{-2})$ s is much smaller than the typical droplet lifetime, which is $\sim O(10^2)$ s. Since the Bond number of the droplet $\mathrm{Bo}=\tfrac{\rho_\ell g R^{2}}{\gamma}\ll1$, where $g$ is the gravitational acceleration and $\gamma$ the surface tension, the influence of gravity on the droplet shape is negligible. 
Droplets evaporate from the BSA coated glass with a pinned contact line, and with a small initial contact angle ($\theta_0\approx 50^\circ$). In this case, the height-averaged radial velocity is given by \cite{marin2011order}
\begin{equation}
   u(r,t) = \frac{2R^2U_e}{\pi  r} \frac{1}{1-t/t_f} \left[\frac{1}{\sqrt{R^{2}-r^{2}}}-\frac{1}{R^{3}} ( R^{2}-r^{2} ) \right].
   \label{eq:capflow}
\end{equation}
We experimentally verified that this velocity field is in good agreement with the measured background flow close to the contact line (see \emph{SI Appendix} S1 B), where Marangoni circulation is negligible compared to the evaporation-driven radially outward flow \cite{Gelderblom2012}. 

We non-dimensionalize Eqs.~(\ref{eq:n*},\ref{eq:capflow}) using
\begin{align}
    \hat{r}&=r/R,     &  \hat{h}&=2h/(\theta_0R), \nonumber \\ 
    \hat{t}&=t/t_f,     &  \hat{u}&=u/U_e,     \label{eq:scaling} \\
    \hat{n}&=n/n_i,      \nonumber
\end{align}
where $t_f$ the total droplet lifetime, $U_e$ is the characteristic capillary flow velocity \cite{Gelderblom2022}, $n_i$ the initial number density of the bacteria, and $h$ the droplet height.
From now on we work with dimensionless quantities $0<\hat{r}<1$ and $0<\hat{t}<1$, such that we can rewrite Eq.~(\ref{eq:n*}) as
\begin{equation}
    \frac{8}{\pi} \frac{\partial}{\partial \hat{t}}\left(\hat{h}\hat{n}\right) + \frac{1}{\hat{r}}\frac{\partial}{\partial \hat{r}}\left[\hat{r}\hat{h}\hat{n}\hat{u}-\frac{1}{\mathrm{Pe}}\hat{r}\hat{h}\frac{\partial \hat{n}}{\partial \hat{r}}\right]=0,
    \label{eq:n_dm}
\end{equation}
with Pe given by (\ref{eq:pec}).
Moore et al.~\cite{moore2021nascent} derived a matched asymptotic solution to Eq.\,(\ref{eq:n_dm}) for the distribution of solute inside an evaporating sessile droplet with a pinned contact line in the limit of large $\mathrm{Pe}$ that is given by 
\begin{equation}
    \hat{n}(\hat{r},\hat{t};\mathrm{Pe})=\frac{m(\hat{r},\hat{t})}{\hat{h}(\hat{r},\hat{t})}=\frac{1}{\hat{h}(\hat{r},\hat{t})}\left[m_0(\hat{r},\hat{t})+\mathrm{Pe}^2M_0\left(\hat{r},\hat{t};\mathrm{Pe}\right)\right],
    \label{eq:n}
\end{equation}
where $m=\hat{h}\hat{n}$ is the amount of solute per unit area, and 
\begin{align}
    \hat{h}(\hat{r},\hat{t})&=(1-\hat{r}^2)(1-\hat{t}),\label{eq:hf}\\
    m_0(\hat{r},\hat{t})&=\left(1-\hat{r}^2\right)^{1/2}(1-\hat{t})^{3/4}\left\{1-\right.\nonumber\\
    &\left.(1-\hat{t})^{3/4}\left[1-(1-\hat{r}^2)^{3/2}\right]\right\}^{1/3},\label{eq:m0}\\
    M_0(\hat{r},\hat{t};\mathrm{Pe})&=\mathrm{Pe}^2(1-\hat{r}) F(\hat{t})\exp{\left(-\mathrm{Pe}\frac{2\sqrt{2}}{\pi}\frac{\left(1-\hat{r}\right)^{1/2}}{1-\hat{t}}\right)},\\
    F(\hat{t})&=\frac{16}{3\pi^4}\frac{1}{(1-\hat{t})^4}\left\{\frac{1}{4}\left[1-(1-\hat{t})^{3/4}\right]^{4/3}\right\},
\end{align}
with $m_0$ the solution in the bulk of the droplet, where mass transport is dominated by advection, and $M_0$ is the solution for the boundary layer at the contact line, where diffusive transport plays a role.

\begin{acknowledgements}The authors thank Ad Holten, Freek Uittert and J{\o}rgen van der Veen for technical support, Janne-Mieke Meijer and Max Schelling for the synthesis of the nm-PEG coated particles, and Alexander Morozov, Suriya Prakash, Anton Darhuber, Tess Homan and Alvaro Mar\'in for fruitful discussions.
HG acknowledges financial support from the Netherlands Organisation for Scientific Research (NWO) through Veni Grant No.~680-47-451.
\end{acknowledgements}


\bibliography{bactbib}

@article{Burriel:2024,
  title={Active Density Pattern Formation in Bacterial Binary Mixtures},
  author={Espada Buriel, S. and Colin, R.},
  journal={Phys. Rev. X. Life},
volume={2},
number={023002},
  year={2024}
}

@article{Sefiane:2010,
  title={On the formation of regular patterns from drying droplets and their potential use for biomedical applications},
  author={Sefiane, K.},
  journal={J. Bion. Eng.},
volume={7},
number={4},
pages={pp.~S82--S93},
  year={2010}
}

@article{Baskaran:2009,
  title={Statistical mechanics and hydrodynamics of bacterial suspensions},
  author={Baskaran, A. and Marchetti, M. C.},
  journal={Proc. Natl Acad. Sci. USA},
volume={106},
pages={15567–15572},
  year={2009}
}

@article{Ezhilan:2013,
  title={Instabilities and nonlinear dynamics of concentrated active suspensions},
  author={Ezhilan, B. and Shelley, M. J. and Saintillan, D. },
  journal={Phys. Fluids},
volume={25},
number={070607},
  year={2013}
}

@article{Trantum:2012,
  title={Biomarker-mediated disruption of coffee-ring formation as a low resource diagnostic indicator},
  author={Trantum, J.R. and Wright, D.W. and Haselton, F.R.},
  journal={Langmuir},
volume={28},
number={4},
pages={pp.~2187--2193},
  year={2012}
}

@article{Deegan:1997,
  title={Capillary flow as the cause of ring stains from dried liquid drops},
  author={Deegan, R.D. and Bakajin, O. and Dupont, T.F. and Huber, G. and Nagel, S.R. and Witten, T.A.},
  journal={Nature},
  volume={389},
  pages={827},
  year={1997},
  publisher={Nature Publishing Group UK London}
}

@article{Stenhammer:2017,
  title={Role of Correlations in the Collective Behavior of Microswimmer Suspensions},
  author={Stenhammar, J. and Nardini, C. and Nash, R.W. and Maren-duzzo, D. and Morozov, A.},
  journal={Phys. Rev. Lett.},
volume={119},
number={028005},
  year={2017}
}

@article{Poulain:2018,
  title={Biosurfactants change the thinning of contaminated bubbles at bacteria-laden
water interfaces},
  author={Poulain, S. and Bourouiba, L.},
  journal={Phys. Rev. Lett.},
volume={121},
number={204502},
  year={2018}
}

@article{Angelini:2009,
  title={Bacillus Subtilis spreads by surfing waves of
surfactant},
  author={Angelini, T. and Roper, M. and Kolter, R. and Weitz, D. and Brenner, M.},
  journal={Proc. Natl. Acad. Sci. USA},
  volume={106},
number={43},
  pages={},
  year={2009}
}

@article{Orevi:2021,
  title={Life in a droplet: microbial ecology in microscopic surface wetness},
  author={Orevi, T. and Kashtan, N.},
  journal={Front. Microbiol.},
  volume={12},
  number={655459},
  year={2021}
}

@article{Grinberg:2019,
  title={Bacterial survival in microscopic wetness},
  author={Grinberg, M. and Orevi, T. and Steinberg, S. and Kashtan, N.},
  journal={eLife},
  volume={8},
  number={e48508},
  year={2019}
}

@article{Knowlton:2018,
  title={Bioaerosol concentrations generated from toilet flushing in a hospital-based patient care setting},
  author={Knowlton, S.D. and Boles, C.L. and Perencevich, E.N. and Diekema, D.J. and Nonnenmann, M.W.},
  journal={Antimicrob. Resist. Infect. Control},
  volume={7},
 number={16},
  year={2018}
}

@article{Gilet:2015,
  title={Fluid fragmentation shapes rain-induced foliar disease transmission},
  author={Gilet, T. and Bourouiba, L.},
  journal={J. R. Soc.},
  volume={12},
 number={104},
  year={2015}
}

@article{Leveau:2019,
  title={A brief from the leaf: latest research to inform our understanding of the phyllosphere microbiome},
  author={Leveau, J.H.J.},
  journal={Curr. Opin. in Microbiol.},
  volume={49},
  pages={41--49},
  year={2019}
}

@article{Colin:2019,
  title={Chemotactic behaviour of Escherichia coli at high cell density},
  author={Colin, R. and Drescher, K. and Sourjik, V.},
  journal={Nat. Commun.},
  volume={10},
  number={5329},
  year={2019}
}

@article{Xu:2023,
  title={Geometrical control of interface patterning underlies active matter invasion},
  author={Xu, H. and Nejad, M.R. and Yeomans, J.M. and Wu, Y.},
  journal={Proc. Natl. Acad. Sci. USA},
  volume={120},
number={30},
  pages={},
  year={2023}
}

@article{Dunkel:2013,
  title={Fluid Dynamics of Bacterial Turbulence},
  author={Dunkel, J. and Heidenreich, S. and Drescher, K. and Wensink, H.H. and B\:ar, M. and Goldstein, R.E.},
  journal={Phys. Rev. Lett.},
volume={110},
number={228102},
  year={2015}
}

@article{Sokolov:2007,
  title={Concentration dependence of the collective dynamics of swimming bacteria},
  author={Sokolov, A. and Aranson, I.S. and Kessler, J.O. and Goldstein, R.E},
  journal={Phys. Rev. Lett.},
volume={98},
number={158102},
  year={2007}
}

@article{Wioland:2016,
 title={Directed collective motion of bacteria under channel confinement},
  author={Wioland, H. and Lushi, E. and Goldstein, R.E},
  journal={New J. Phys.},
volume={18},
number={075002},
  year={2016}
}

@article{Skultety:2020,
  title={Swimming suppresses correlations in dilute suspensions of pusher microorganisms},
  author={Skult\'ety, V. and Nardini, C. and Stenhammer, J. and Marenduzzo, D. and Morozov, A.},
  journal={Phys. Rev. X},
  volume={10},
  pages={031059 },
  year={2020}
}

@article{Schofield:2018,
  title={The lifetimes of evaporating sessile droplets are significantly extended by strong thermal effects},
  author={Schofield, F.G.H. and Wilson, S.K. and Pritchard, D. and Sefiane, K.},
  journal={J. Fluid Mech.},
  volume={851},
  pages={231--244},
  year={2018}
}

@article{Boulogne:2019,
 author = {Boulogne, F.},
 journal = {Eur. Phys. J. E},
 number = {4},
 pages = {51},
 title = {Cheap and versatile humidity regulator for environmentally controlled experiments},
 volume = {42},
 year = {2019}
}

@article{berke2008,
  title={Hydrodynamic attraction of swimming microorganisms by surfaces},
  author={Berke, Allison P and Turner, Linda and Berg, Howard C and Lauga, Eric},
  journal={Phys. Rev. Lett.},
  volume={101},
  number={3},
  pages={038102},
  year={2008},
  publisher={APS}
}

@article{Kasyap2014,
  doi = {10.1063/1.4901958},
  url = {https://doi.org/10.1063/1.4901958},
  year = {2014},
  month = nov,
  publisher = {{AIP} Publishing},
  volume = {26},
  number = {11},
  pages = {111703},
  author = {T.V. Kasyap and D.L. Koch and M. Wu},
  title = {Bacterial collective motion near the contact line of an evaporating sessile drop},
  journal = {Phys. Fluids}
}

@article{kodger2015precise,
  title={Precise colloids with tunable interactions for confocal microscopy},
  author={Kodger, T.E. and Guerra, R.E. and Sprakel, J.},
  journal={Sci. Rep.},
  volume={5},
  number={1},
  pages={14635},
  year={2015},
  publisher={Nature Publishing Group UK London}
}

@article{kasyap2012chemotaxis,
  title={Chemotaxis driven instability of a confined bacterial suspension},
  author={Kasyap, TV and Koch, Donald L},
  journal={Phys. Rev. Lett.},
  volume={108},
  number={3},
  pages={038101},
  year={2012},
  publisher={APS}
}

@article{barnkob2021,
  title={DefocusTracker: A Modular Toolbox for Defocusing-based, Single-Camera, 3D Particle Tracking},
  author={Barnkob, Rune and Rossi, Massimiliano},
journal={J. Open Res. Softw.},
volume={9},
number={1},
  year={2021},
  publisher={Ubiquity Press}
}

@article{bialopiotrowicz2001,
  title={Wettability and surface free energy of bovine serum albumin films},
  author={Bia{\l}opiotrowicz, Tomasz and Ja{\'n}czuk, Bronis{\l}aw},
  journal={J. Surfactants Deterg.},
  volume={4},
  pages={287--292},
  year={2001},
  publisher={Springer}
}

@article{bourouiba2021,
  title={The fluid dynamics of disease transmission},
  author={Bourouiba, Lydia},
  journal={Ann. Rev. Fluid Mech.},
  volume={53},
  pages={473--508},
  year={2021},
  publisher={Annual Reviews}
}

@article{thokchom2014,
  title={Fluid flow and particle dynamics inside an evaporating droplet containing live bacteria displaying chemotaxis},
  author={Thokchom, Ashish Kumar and Swaminathan, Rajaram and Singh, Anugrah},
  journal={Langmuir},
  volume={30},
  number={41},
  pages={12144--12153},
  year={2014},
  publisher={ACS Publications}
}

@article{turner2010visualization,
  title={Visualization of flagella during bacterial swarming},
  author={Turner, Linda and Zhang, Rongjing and Darnton, Nicholas C and Berg, Howard C},
  journal={J. Bacteriol.},
  volume={192},
  number={13},
  pages={3259--3267},
  year={2010},
  publisher={Am Soc Microbiol}
}

@article{turner2012growth,
  title={Growth of flagellar filaments of Escherichia coli is independent of filament length},
  author={Turner, Linda and Stern, Alan S and Berg, Howard C},
  journal={J. Bacteriol.},
  volume={194},
  number={10},
  pages={2437--2442},
  year={2012},
  publisher={Am Soc Microbiol}
}

@article{tuval2005,
  title={Bacterial swimming and oxygen transport near contact lines},
  author={Tuval, I. and Cisneros, L. and Dombrowski, C. and Wolgemuth, C.W. and Kessler, J.O. and Goldstein, R.E.},
  journal={Proc. Natl. Acad. Sci.},
  volume={102},
  number={7},
  pages={2277--2282},
  year={2005},
  publisher={National Acad Sciences}
}

@article{sempels2013,
  title={Auto-production of biosurfactants reverses the coffee ring effect in a bacterial system},
  author={Sempels, Wouter and De Dier, Raf and Mizuno, Hideaki and Hofkens, Johan and Vermant, Jan},
  journal={Nat. Commun.},
  volume={4},
  number={1},
  pages={1757},
  year={2013},
  publisher={Nature Publishing Group UK London}
}

@article{nellimoottil2007,
  title={Evaporation-induced patterns from droplets containing motile and nonmotile bacteria},
  author={Nellimoottil, Tittu Thomas and Rao, Pinjala Nagaraju and Ghosh, Siddhartha Sankar and Chattopadhyay, Arun},
  journal={Langmuir},
  volume={23},
  number={17},
  pages={8655--8658},
  year={2007},
  publisher={ACS Publications}
}

@article{moore2021nascent,
  title={The nascent coffee ring: how solute diffusion counters advection},
  author={Moore, Matthew R and Vella, Dominic and Oliver, James M},
  journal={J. Fluid Mech.},
  volume={920},
  pages={A54},
  year={2021},
  publisher={Cambridge University Press}
}

@article{Andac2019,
  doi = {10.1039/c8sm01350k},
  url = {https://doi.org/10.1039/c8sm01350k},
  year = {2019},
  publisher = {Royal Society of Chemistry ({RSC})},
  volume = {15},
  number = {7},
  pages = {1488--1496},
  author = {T. Andac and P. Weigmann and S.K.P. Velu and others},
  title = {Active matter alters the growth dynamics of coffee rings},
  journal = {Soft Matter}
}

@article{Gelderblom2012,
  title={Stokes flow near the contact line of an evaporating drop},
  author={Gelderblom, Hanneke and Bloemen, Oscar and Snoeijer, Jacco H.},
  journal={J. Fluid Mech.},
  volume={709},
  pages={69--84},
  year={2012},
  publisher={Cambridge University Press}
}

@article{Dombrowski2004,
  doi = {10.1103/physrevlett.93.098103},
  url = {https://doi.org/10.1103/physrevlett.93.098103},
  year = {2004},
  month = aug,
  publisher = {American Physical Society ({APS})},
  volume = {93},
  number = {9},
  author = {Dombrowski, C. and Cisneros, L. and Chatkaew, S. and Goldstein, R.E. and Kessler, J.},
  title = {Self-Concentration and Large-Scale Coherence in Bacterial Dynamics},
  journal = {Phys. Rev. Lett.}
}

@article{Gelderblom2022,
  doi = {10.1039/d2sm00931e},
  url = {https://doi.org/10.1039/d2sm00931e},
  year = {2022},
  publisher = {Royal Society of Chemistry ({RSC})},
  volume = {18},
  number = {45},
  pages = {8535--8553},
  author = {H. Gelderblom and C. Diddens and A. Marin},
  title = {Evaporation-driven liquid flow in sessile droplets},
  journal = {Soft Matter}
}

@article{SchwarzLinek2016,
  doi = {10.1016/j.colsurfb.2015.07.048},
  url = {https://doi.org/10.1016/j.colsurfb.2015.07.048},
  year = {2016},
  month = jan,
  publisher = {Elsevier {BV}},
  volume = {137},
  pages = {2--16},
  author = {J. Schwarz-Linek and J. Arlt and A. Jepson and others},
  title = {\textit{Escherichia coli} as a model active colloid: A practical introduction},
  journal = {Colloids Surf. B}
}

@article{altendorf2009osmotic,
  title={Osmotic stress},
  author={Altendorf, K. and Booth, I.R. and Gralla, J.A.Y. and Greie, J.-C. and Rosenthal, A.Z. and Wood, J.M.},
  journal={EcoSal Plus},
  volume={3},
  number={2},
  pages={10--1128},
  year={2009},
  publisher={Am.Soc Microbiol}
}

@article{dugyala2015evaporation,
  title={Evaporation of sessile drops containing colloidal rods: coffee-ring and order--disorder transition},
  author={Dugyala, Venkateshwar Rao and Basavaraj, Madivala G},
  journal={J. Phys. Chem. B},
  volume={119},
  number={9},
  pages={3860--3867},
  year={2015},
  publisher={ACS Publications}
}

@article{shahidzadeh-bonn_rafaï_azouni_bonn_2006, title={Evaporating droplets}, 
volume={549}, 
journal={J. Fluid Mech.}, 
publisher={Cambridge University Press}, author={Shahidzaheh-Bonn, N. and Rafa\:i, S. and Azouni, A. and Bonn, D.}, 
year={2006}, 
pages={307–313}}

@article{thery2020self,
  title={Self-organisation and convection of confined magnetotactic bacteria},
  author={Th{\'e}ry, A. and Le Nagard, L. and Ono-dit-Biot, J.-C. and Fradin, C. and Dalnoki-Veress, K. and Lauga, E.},
  journal={Sci. Rep.},
  volume={10},
  number={1},
  pages={13578},
  year={2020},
  publisher={Nature Publishing Group UK London}
}

@article{wioland2013confinement,
  title={Confinement stabilizes a bacterial suspension into a spiral vortex},
  author={Wioland, Hugo and Woodhouse, Francis G and Dunkel, J{\"o}rn and Kessler, John O and Goldstein, Raymond E},
  journal={Phys. Rev. Lett.},
  volume={110},
  number={26},
  pages={268102},
  year={2013},
  publisher={APS}
}

@article{lushi2014fluid,
  title={Fluid flows created by swimming bacteria drive self-organization in confined suspensions},
  author={Lushi, Enkeleida and Wioland, Hugo and Goldstein, Raymond E},
  journal={Proc. Natl. Acad. Sci.},
  volume={111},
  number={27},
  pages={9733--9738},
  year={2014},
  publisher={National Acad Sciences}
}

@article{drescher2009dancing,
  title={Dancing volvox: hydrodynamic bound states of swimming algae},
  author={Drescher, Knut and Leptos, Kyriacos C and Tuval, Idan and Ishikawa, Takuji and Pedley, Timothy J and Goldstein, Raymond E},
  journal={Phys. Rev. Lett.},
  volume={102},
  number={16},
  pages={168101},
  year={2009},
  publisher={APS}
}

@article{marin2011order,
  title={Order-to-disorder transition in ring-shaped colloidal stains},
  author={Marin, Alvaro G and Gelderblom, Hanneke and Lohse, Detlef and Snoeijer, Jacco H},
  journal={Phys. Rev. Lett.},
  volume={107},
  number={8},
  pages={085502},
  year={2011},
  publisher={APS}
}

@article{Marin2019,
  doi = {10.1103/physrevfluids.4.041601},
  url = {https://doi.org/10.1103/physrevfluids.4.041601},
  year = {2019},
  month = apr,
  publisher = {American Physical Society ({APS})},
  volume = {4},
  number = {4},
  author = {A. Marin and S. Karpitschka and D. Noguera-Mar{\'{\i}}n and others},
  title = {Solutal Marangoni flow as the cause of ring stains from drying salty colloidal drops},
  journal = {Phys. Rev. Fluids}
}

@article{hyon2012wiggling,
  title={The wiggling trajectories of bacteria},
  author={Hyon, Yunkyong and Powers, Thomas R and Stocker, Roman and Fu, Henry C and others},
  journal={J. Fluid Mech.},
  volume={705},
  pages={58--76},
  year={2012},
  publisher={Cambridge University Press}
}

@article{seyfert2021evaporation,
  title={Evaporation-driven colloidal cluster assembly using droplets on superhydrophobic fractal-like structures},
  author={Seyfert, Carola and Berenschot, Erwin JW and Tas, Niels R and Susarrey-Arce, Arturo and Marin, Alvaro},
  journal={Soft matter},
  volume={17},
  number={3},
  pages={506--515},
  year={2021},
  publisher={Royal Society of Chemistry}
}

@article{drescher2011fluid,
  title={Fluid dynamics and noise in bacterial cell--cell and cell--surface scattering},
  author={Drescher, Knut and Dunkel, J{\"o}rn and Cisneros, Luis H and Ganguly, Sujoy and Goldstein, Raymond E},
  journal={Proc. Natl. Acad. Sci.},
  volume={108},
  number={27},
  pages={10940--10945},
  year={2011},
  publisher={National Acad Sciences}
}

@article{pierce2018hydrodynamic,
  title={Hydrodynamic interactions, hidden order, and emergent collective behavior in an active bacterial suspension},
  author={Pierce, CJ and Wijesinghe, H and Mumper, E and Lower, BH and Lower, SK and Sooryakumar, R},
  journal={Phys. Rev. Lett.},
  volume={121},
  number={18},
  pages={188001},
  year={2018},
  publisher={APS}
}

@book{Lauga2020,
  doi = {10.1017/9781316796047},
  url = {https://doi.org/10.1017/9781316796047},
  year = {2020},
  month = sep,
  publisher = {Cambridge University Press},
  author = {E. Lauga},
  title = {The Fluid Dynamics of Cell Motility}
}

@article{majee2021,
  title={Spatiotemporal evaporating droplet dynamics on fomites enhances long term bacterial pathogenesis},
  author={Majee, Sreeparna and Chowdhury, Atish Roy and Pinto, Roven and Chattopadhyay, Ankur and Agharkar, Amey Nitin and Chakravortty, Dipshikha and Basu, Saptarshi},
  journal={Communications Biology},
  volume={4},
  number={1},
  pages={1173},
  year={2021},
  publisher={Nature Publishing Group UK London}
}

@article{richard2020,
  title={Hydrophobicity of abiotic surfaces governs droplets deposition and evaporation patterns},
  author={Richard, Elodie and Dubois, Thomas and Allion-Maurer, Audrey and Jha, Piyush Kumar and Faille, Christine},
  journal={Food microbiology},
  volume={91},
  pages={103538},
  year={2020},
  publisher={Elsevier}
}

@article{ma2022crack,
  title={Crack patterns of drying dense bacterial suspensions},
  author={Ma, Xiaolei and Liu, Zhengyang and Zeng, Wei and Lin, Tianyi and Tian, Xin and Cheng, Xiang},
  journal={Soft Matter},
  volume={18},
  number={28},
  pages={5239--5248},
  year={2022},
  publisher={Royal Society of Chemistry}
}

@article{hennes2017active,
  title={Active depinning of bacterial droplets: The collective surfing of Bacillus subtilis},
  author={Hennes, M and Tailleur, J and Charron, G and Daerr, A},
  journal={Proc. Natl. Acad. Sci.},
  volume={114},
  number={23},
  pages={5958--5963},
  year={2017},
  publisher={National Acad Sciences}
}

\end{document}